\documentclass[preprint,11]{aastex}
\begin{document}
\title{ACCRETION THROUGH THE INNER EDGES OF PROTOPLANETARY DISKS BY A GIANT SOLID STATE PUMP}
\author{T. Kelling and G. Wurm}
\affil{Faculty of Physics, University of Duisburg-Essen, Lotharstr. 1, 47048 Duisburg, Germany}
\email{thorben.kelling@uni-due.de\\The Astrophysical Journal Letters, accepted 2013 July 26,}

\received{2013 June 5}
\accepted{2013 July 26}

\begin{abstract}
At the inner edge of a protoplanetary disk solids are illuminated by stellar light. This illumination 
heats the solids and creates temperature gradients along their surfaces. 
Interactions with ambient gas molecules lead to a radial net gas flow. Every illuminated solid particle within the edge is an individual small gas pump transporting gas inward. 
In total the inner edge can provide local mass flow rates as high as $\dot{M} = 10^{-5} M_{\odot}$ yr$^{-1}$.
\end{abstract}

\maketitle

\section{Introduction}

The accretion of gas by a young star by means of an accretion disk (called a protoplanetary disk in the connection to planet formation) is fundamental to its evolution. Typical accretion rates of young stars are on the order of $10^{-6} M_{\odot}$ yr$^{-1}$ decreasing with the age of the disk \citep{armitage2003,manara2012}. 

The inner edge of a protoplanetary disk is unique with respect to accretion 
as matter is transported from a high-density outer disk ({optically thick}) to a depleted inner region (optically thin).
Specific to the edge is the {direct} illumination {of solids} by stellar light.
In recent years, the motion of small particles at the edge by photophoresis was considered, assuming the solids to be ''test particles'', with no feed-back of the particle motion to the gas \citep{haack2007, loesche2012, wurm2013}.
As an extreme, the whole inner edge of the disk might move 
outward \citep{krauss2007, moudens2011}. However, if all particles at the edge are subject to photophoretic forces, the response of the gas can no longer be neglected.
In fact, on one side, photophoretic forces push solids away from a light source (star).
On the other side, focusing attention on the gas, photophoresis is based on a motion of gas molecules towards the light source (star) and every illuminated particle acts like
a small micro pump, pushing gas radially inward.

\section{Thermal gradient pumps}

Temperature gradients on a solid surface induce gas flows and particle motion. A prominent example of particle motion is the light mill \citep{crookes1874}. A less known but nonetheless impressive example of gas flow is the Knudsen compressor. Tubes with a cross section of the order of the mean free path of the gas molecules heated at one end allowed \cite{knudsen1909} to compress a gas by a factor of about 10 \citep{knudsen1909}. We consider temperature gradients induced by absorbing a light flux $I$ at an inner edge of a protoplanetary disk here, varying with distance $d$ as $I=I_0\left( \frac{d[\mathrm{AU}]}{1 \mathrm{AU}}\right)^{-2}$ with $I_0=1360$ W m$^{-2}$. The absorption leads to a small temperature difference $\Delta T = T_w-T_c$ on a particle with radius $r$ with $T_w$ and $T_c$ as the temperatures at the illuminated warm front side and cold back side, respectively. Assuming perfect absorption at the front side it is \citep{rohatschek1995} $\Delta T = I  r  \kappa_p^{-1}$ with the thermal conductivity of the particle taken to be $\kappa_p=0.1$ W/(m K). Gas molecules at ambient average temperature $T$ diffusely scattered from a local surface for $Kn>1$, where $Kn$ is the Knudsen number, acquire the local surface temperature $T_w$ or $T_c$. This induces a photophoretic force on the particle which accelerates the particle. In equilibrium with the gas drag force the final velocity of the particle is \citep{kelling2011}
\begin{equation}
v_p=\frac{\alpha_{acc}\epsilon J_1 I r}{3\kappa_p\sqrt{8T/\pi}}\sqrt{\frac{R}{M}}.\label{eq:driftp}
\end{equation}
It is $\alpha_{acc}=1$ the accommodation coefficient of the gas molecules, $\epsilon=0.7$ an empirical factor \citep{blum1996}, $\left| J_1\right|=0.5$ the asymmetry factor (internal heat sources of the illuminated particle), $R=8.3$ J mol$^{-1}$K$^{-1}$ the universal gas constant, $r$ the particle radius and $M=2.34\times 10^{-3}$ kg mol$^{-1}$ the molar mass of the gas.

The gas in the disk moves inward with the particles velocity times the fraction of dust $f_p\leq 1$ (momentum balance), i.e. $v_{gas}=f_pv_p$. At the inner edge of the disk the stellar radiation is completely absorbed. If the transition is filled with particles of one size $r$, every particle acts as a micro pump (Figure 1).
\begin{figure}
 \centering
\includegraphics[width=0.7\textwidth]{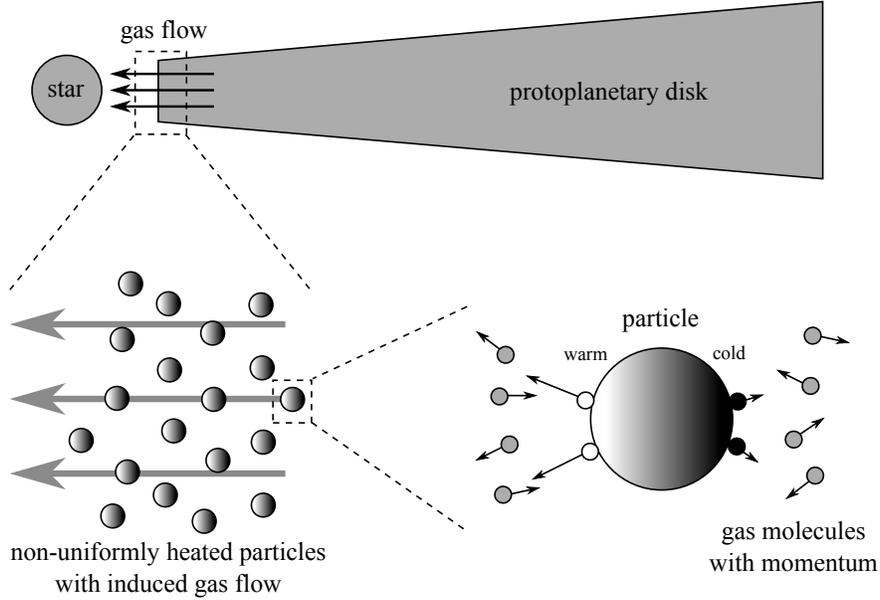}
 \caption{Illustration of an inner edge acting as a giant solid state pump.}
\end{figure}
The total gas mass flow is then
\begin{equation}
\dot{M}=\Sigma\cdot v_{gas}\cdot 2\pi d,
\end{equation}
with $\Sigma$ as the surface density. To quantify this we assume a minimum mass solar nebula \citep{hayashi1985}  
with $\Sigma=1.7\times 10^3 \left( \frac{d[\mathrm{AU}]}{1 \mathrm{AU}}\right)^{-3/2}$ g cm$^{-2}$, $T = T_0 \left( \frac{d[\mathrm{AU}]}{1 \mathrm{AU}}\right)^{-1 / 2}$ and $T_0=280$ K. This leads to a total mass flow per year
\begin{equation}
\dot{M} = f_p\xi r\left( \frac{d[\mathrm{AU}]}{1 \mathrm{AU}}\right)^{-9/4},
\end{equation}
where the parameter $\xi = 8.97\times 10^{-4}$ $M_{\odot}$ yr$^{-1}$ m$^{-1}$ contains all gas and particle properties. In Figure 2 we show the dependency of the giant solid state pump on the position of the inner edge in the protoplanetary disk, the particle size and the fraction of dust $f_p=1/100$ (solid lines) and $f_p=1$ (small dashed lines). In addition, we compare the gas mass flow rates with accretion rates of turbulent disks (large dashed lines) with $\dot{M}_{turb}=3\pi \Sigma \alpha c_s^2/\Omega$ and $\alpha$ as turbulence parameter, $c_s=\sqrt{RT/M}$ as sound speed and $\Omega$ as Keplerian angular velocity \citep{jacquet2013}.
\begin{figure}
 \centering
\includegraphics[width=0.7\textwidth]{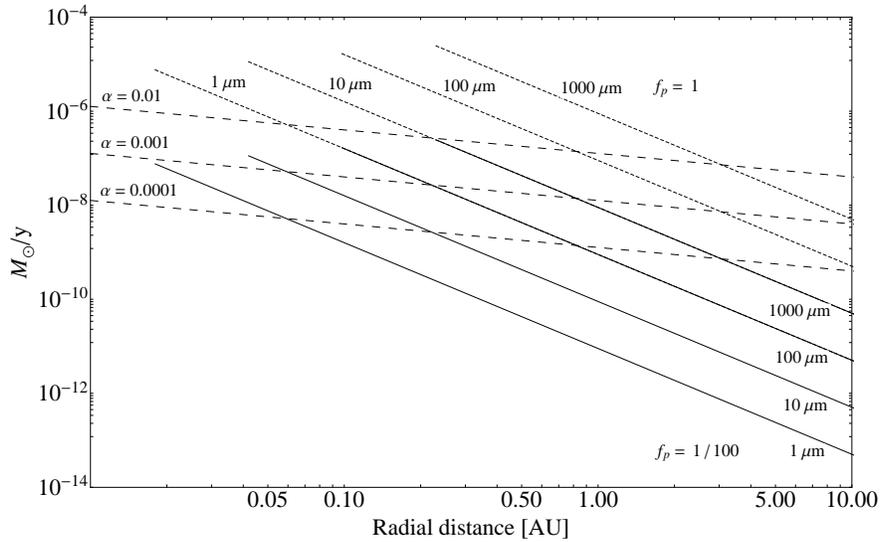}
 \caption{Mass flow rate in a minimum mass solar nebula for $Kn>1$. The lines for the different particle sizes end at $Kn=1$ where the hydrodynamic regime changes \citep{takeuchi2008}. The large dashed lines indicate the accretion rate of turbulent discs for different $\alpha$ parameters, the small dashed lines are mass flow rates for a dust fraction of 1 and the solid lines are for a dust fraction of 1/100.}
\end{figure}

\section{Conclusions}

The mass flow rates through the inner edge of a minimum mass solar nebula estimated above can be as high as $\dot{M}=10^{-5}$ $M_{\odot}$ yr$^{-1}$ for dust fractions of 1 (Figure 2). The mass flow rates calculated are beyond accretion rates of T-Tauri disks close to the star and fall below these at some AU distance but the solid state pumps act only locally at the inner edge. 

If the general viscous evolution were very slow (no gas accretion), the inner edge would just drain a reservoir on the dark side and increase the ambient pressure on the bright side. As the first phases of planet formation are tied to gas and particle concentrations this might
influence local formation of larger bodies in the disks \citep{chiang2010,blum2008}.

The giant solid state pump might therefore be a 
significant active mechanism in the hydrodynamics at the inner edge of protoplanetary disks influencing the disk's evolutions and planet formation.

%\begin{thebibliography}{1}
%\bibitem{blum2008} Blum, J. {\&} Wurm, G., ARA{\&}A, 46, 21
%\bibitem{chiang2010} Chiang, E. {\&} Youdin, A. N.,  Annual Review of Earth and Planetary Sciences, 38, 493
%\bibitem{crookes1874} Crookes, W., Proceedings of the Physical Society of London, 1,
%35
%\bibitem{dominik2011} Dominik, C. {\&}  Dullemond, C. P. 2011, A{\&} A, 531, A101
%\bibitem{haack2007} Haack, H. {\&}  Wurm, G. 2007, Meteoritics and Planetary Science
%Supplement, 42, 5157
%\bibitem{hayashi1985} Hayashi, C., Nakazawa, K., {\&} Nakagawa, Y. 1985, in Protostars and
%Planets II, ed. D. C. Black {\&} M. S. Matthews, 1100–1153
%\bibitem{knudsen1909} Knudsen, M. 1909, Annalen der Physik, 336, 633
%\bibitem{krauss2007} Krauss, O., Wurm, G., Mousis, O., et al. 2007, A{\&}A, 462, 977
%\bibitem{Loesche2012} Loesche, C. {\&} Wurm, G. 2012, A{\&}A, 545, A36
%\bibitem{manara2012} Manara, C. F., Robberto, M., Da Rio, N., et al. 2012, ApJ, 755, 154
%\bibitem{moudens2011} Moudens, A., Mousis, O., Petit, J.-M., et al. 2011, A{\&}A, 531, 106
%\bibitem{rohatschek1995} Rohatschek, H. 1995, Journal of Aerosol Science, 26, 717
%\bibitem{wurm2013} Wurm, G., Trieloff, M., {\&} Rauer, H. 2013, ApJ, 769, 78
%\end{thebibliography}

%\section{References}
%
%\bibliography{small2} 
%%\bibliography{small}
%\bibliographystyle{aa}
\end{document}